\documentclass[twoside,a4paper,11pt]{sea10}
\usepackage{graphicx}
\begin{document}
\pagenumbering{arabic}
\pagestyle{myheadings}
\thispagestyle{empty}
{\flushleft\includegraphics[width=\textwidth,bb=58 650 590 680]{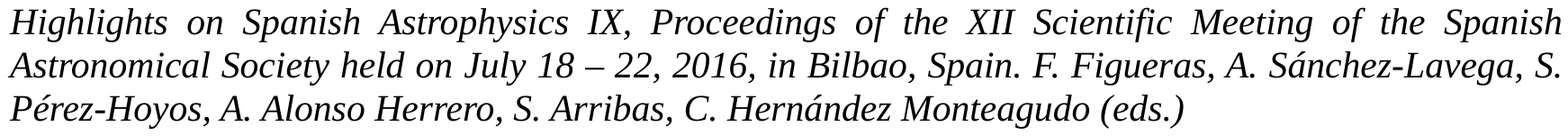}}
\vspace*{0.2cm}
\begin{flushleft}
{\bf {\LARGE
%
The (dark) halo-to-stellar mass ratio in the Spitzer Survey of Stellar Structure in Galaxies (S$^4$G)
%
}\\
\vspace*{1cm}
Sim\'on D\'iaz-Garc\'ia$^1$, 
Heikki Salo$^1$, 
Eija Laurikainen$^1$, 
and 
Ryan Leaman$^2$
%
}\\
\vspace*{0.5cm}
%
$^{1}$
Astronomy Research Group, University of Oulu, FI-90014 Finland, email: {\tt simon.diazgarcia@oulu.fi}
\\
$^{2}$
Max-Planck Institut f\"ur Astronomie, K\"onigstuhl 17, D-69117 Heidelberg, Germany
%
\end{flushleft}
%
\markboth{
The halo-to-stellar mass ratio in the S$^{4}$G
}{ 
%
%
}
\thispagestyle{empty}
\vspace*{0.4cm}
\begin{minipage}[l]{0.09\textwidth}
\ 
\end{minipage}
\begin{minipage}[r]{0.9\textwidth}
\vspace{1cm}
\section*{Abstract}{\small
We use 3.6~$\mu$m photometry for 1154 disk galaxies ($i<65^{\circ}$) in the 
Spitzer Survey of Stellar Structure in Galaxies (S$^{4}$G, \cite{Sheth2010}) to obtain the stellar component of the circular velocity. 
By combining the disk+bulge rotation curves with H{\sc\,i} line width measurements from the literature, 
we estimate the ratio of the halo-to-stellar mass ($M_{\rm halo}/M_{\ast}$) within the optical disk, and compare it to the total stellar mass ($M_{\ast}$). 
We find the $M_{\rm halo}/M_{\ast}-M_{\ast}$ relation in good agreement with the best-fit model at z$\approx$0 
in $\Lambda$CDM cosmological simulations (e.g. \cite{Moster2010}), 
assuming that the dark matter halo within the optical radius comprises a constant fraction ($\sim4\%$) of its total mass.
%
\normalsize}
\end{minipage}
%
%
%
\section{Introduction \label{intro}}

In the lambda cold dark matter ($\Lambda$CDM) cosmological paradigm, galaxies are formed from the cooling and condensation of gas in the center of dark matter halos (\cite{White},\cite{Fall}). 
A challenge for astronomers is to link the properties of present-day galaxies to their parent host halos with the aid of observations and cosmological simulations.

Statistical methods, based on cosmological simulations, to address the interdependence between the dark halo and the stellar mass 
are the halo occupation distribution (e.g. \cite{Peacock}) or the conditional luminosity function (e.g. \cite{vandenBosch}), 
which use large galaxy surveys to calculate the probability distribution of a given halo 
to host certain number of galaxies of particular properties (e.g. total luminosity). 
Similarly, abundance matching (e.g. \cite{Vale}) links the abundances of halos (and subhalos) to those 
of galaxies, under the assumption of a monotonic relation between the dark matter and the stellar mass. 
The stellar-to-halo mass ratio can also be measured directly based on lensing analysis (e.g. \cite{Mandelbaum}) or 
the dynamics of satellite galaxies (e.g. \cite{Klypin}).

From the above techniques, it has been found that the stellar-to-halo mass ratio as a function of halo mass, 
predicted from $\Lambda$CDM cosmological simulations, has a maximum near $M_{\rm halo}\approx10^{12}M_{\odot}$, 
where the star formation efficiency is highest, and drops towards lower- and larger-mass halos. 
This trend is thought to arise from the combined effect of supernova feedback, 
which is more efficient at reheating and dislodging the gas in low-mass halos, and AGN feedback, which is more effective in high mass halos 
(see \cite{Moster2010}, and references therein). 

Combining the disk rotation curves inferred from 3.6~$\mu$m photometry with the H{\sc\,i} velocity measurements from the literature, 
D\'iaz-Garc\'ia et al. (\cite{DG2016b}) obtained an estimate of the halo-to-stellar mass ratio ($M_{\rm halo}/M_{\ast}$) within the optical radius. 
Systems with $T\ge 5$ or $M_{\ast}\le10^{10}M_{\odot}$ were found to be more dark matter dominated inside the optical disk than their 
early-type galaxies (in agreement with \cite{Courteau} and \cite{Falcon}), which most likely affects the disk stability properties. 
Using first-order rotation curve decomposition models (\cite{DG2016b}), which were based on universal rotation curve models, 
we showed that only $\sim 10\%$ of the non-highly inclined galaxies in the S$^4$G were maximal according to the criterion by \cite{Sackett}.

Under the assumption that the halo within the optical disk contributes approximately a constant fraction of the total halo mass ($\sim4\%$), 
D\'iaz-Garc\'ia et al. (\cite{DG2016b}) found that the trend of the $M_{\rm halo}/M_{\ast}-M_{\ast}$ relation agreed with the prediction of $\Lambda$CDM models, 
fitted at $z\approx0$ based on abundance matching and halo occupation distribution methods (e.g. \cite{Moster2010},\cite{Behroozi},\cite{Guo},\cite{Leauthaud}).
%
%
\section{Stellar contribution to the circular velocity}
%
%
Using the polar method in NIR-QB code (\cite{Salo99},\cite{Laurikainen02}), 
we inferred the gravitational potential of 1154 disk galaxies ($i<65^{\circ}$ in S$^4$G Pipeline 4, \cite{Salo15}) 
from the deprojected 3.6~$\mu$m photometric images (\cite{DG2016b}), taken from the S$^4$G sample (\cite{Sheth2010}). 
From the gravitational potential of the galaxies and their mean radial force field, 
we calculated the disk+bulge stellar contribution to the circular velocity as follows:
\begin{equation}
V_{\rm disk}(r)=V_{\rm 3.6\mu m}(r)=\sqrt{\Upsilon_{3.6 \rm \mu m}\left<F_{\rm R}(r)\right> r}, 
\end{equation}
where $r$ is the galactocentric radius, $\rm F_{\rm R}$ corresponds to the radial force obtained for $M/L=1$, 
and $\Upsilon_{3.6 \rm \mu m}=0.53$ is the mass-to-light ratio at 3.6~$\mu$m obtained by \cite{Eskew2012}, 
which is assumed to be constant throughout the disk. 
%
%
\section{Estimate of halo-to-stellar mass ratio}\label{darkmatter}
%
%

We estimate the maximum circular velocities of the galaxies ($V_{\rm HI}^{\rm max}$) from the H{\sc\,i} line widths ($W_{\rm mx}^{\rm av}$), 
available in the Cosmic Flows project (\cite{Courtoisa},\cite{Courtoisb}) and HyperLEDA\footnote{We acknowledge the usage of the database (http://leda.univ-lyon1.fr).}, 
corrected for the disk inclination (from \cite{Salo15}):
\begin{equation}\label{gas_flows}
V_{\rm HI}^{\rm max}=W_{\rm mx}^{\rm av}/(2\,\rm sin\,i).
\end{equation} 

Using $V_{\rm 3.6\mu m}$ and $V_{\rm HI}^{\rm max}$, we obtained a first-order estimate of the halo-to-stellar mass ratio ($M_{\rm h}/M_{\ast}$) within the optical radius ($R_{\rm opt}$). 
$R_{\rm opt}$ is defined as the radius enclosing $83\%$ of the light in the blue band, 
what corresponds to $\sim 3.2h_{\rm R}$ for an exponential disk (e.g. \cite{Swaters}). 
$h_{\rm R}$ is taken from \cite{Salo15}. We assumed that (i) the gas contribution to the rotation curve is modest at the optical radius (e.g. \cite{Verheijen}), 
(ii) the circular velocity at $R_{\rm opt}$ is approximately the observed maximum velocity: 
\begin{equation}\label{dmeqropt}
(V_{\rm HI}^{\rm max})^{2}\approx V_{\rm 3.6\mu m}^{2}(R_{\rm opt})+V_{\rm halo}^{2}(R_{\rm opt}),
\end{equation}
and (iii) the disks are exponential. Based on these assumptions we derive
\begin{equation}
M_{\rm h}/M_{\ast}(<R_{\rm opt})\approx F\cdot\bigg(\frac{(V_{\rm HI}^{\rm max})^{2}}{V_{3.6 \mu \rm m}^{2}(R_{\rm opt})}-1\bigg),
\label{halo-to-stellar-eq}
\end{equation}
where $F$ corresponds to the ratio between the mass contained by a spherical distribution and that enclosed by an exponential disk yielding 
a similar radial force at $R_{\rm opt}$ \cite{Binney}. In the range $2 \le r/h_{\rm R} \le 4$, $F\approx1.34$.
%
%
\section{The $\bf M_{\rm \bf halo}/M_{\bf\ast}-M_{\bf\ast}$ relation}\label{dm_section}
%
%
\begin{figure}
\centering
   \includegraphics[width=0.65\textwidth]{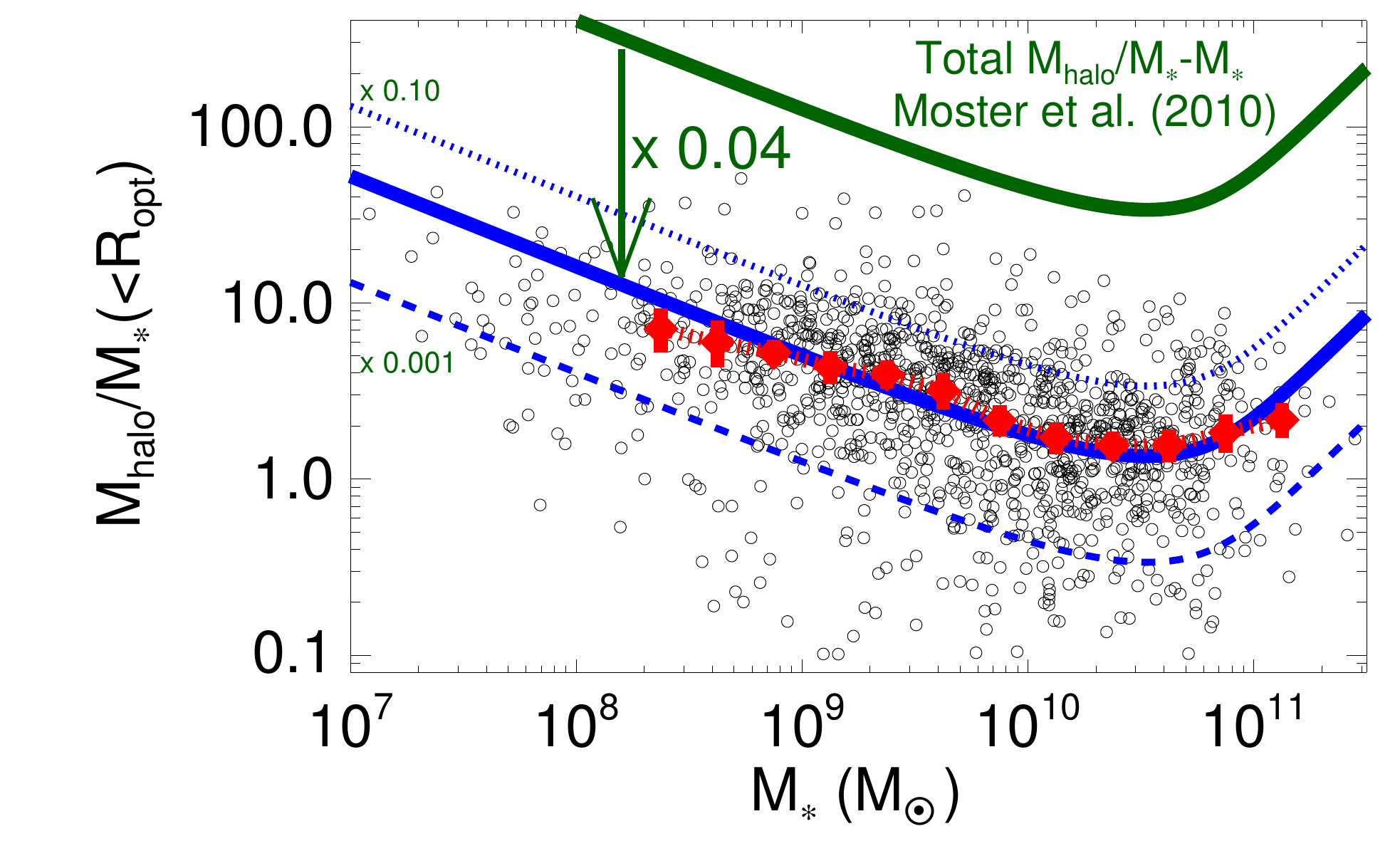}
\vspace{-0.3cm}
\caption{
Halo-to-stellar mass ratio within the optical disk versus total stellar mass (circles, reproduced from \cite{DG2016b}). Red dots represent the running median. 
Lines represent the prediction from cosmological simulations for the 
total halo-to-stellar mass ratio from \cite{Moster2010} (green line), also scaled down by a factor 0.04 (solid blue line), showing agreement with our data. 
Dotted and dashed lines correspond to the same prediction scaled down by factors 0.1 and 0.01, respectively.
}
\label{figuredm}
\end{figure}
%
%
In Fig.~\ref{figuredm} we show the halo-to-stellar mass ratio within the optical radius as a function of total stellar mass ($M_{\ast}$, taken from \cite{Munoz2015}). 
The $M_{\rm halo}/M_{\ast}(<R_{\rm opt})$-$M_{\ast}$ statistical trend agrees with the prediction of the best-fit model at $z\approx0$ 
in the $\Lambda$CDM predictions from cosmological simulations, 
if the models are scaled down by a factor $\sim0.04$ (see Fig.~\ref{figuredm}, and also Fig.~6 in \cite{DG2016b}). 
This means that the amount of dark matter within the optical disk is $\sim 4\%$ that of the total host halo, on average.

Despite the large scatter, the predicted turnover at the high mass end ($M_{\ast}\approx 10^{10.6}M_{\odot}$) in the $\Lambda$CDM models seems to follow 
the observations too, but a more complete sampling of the early-type galaxies is needed. 
Fainter systems ($M_{\ast}\le10^{10}M_{\odot}$) are more dark matter dominated within the optical 
radius than the more massive early-type systems, whose mean $M_{\rm h}/M_{\ast}(<R_{\rm opt})$ is $\sim 2$.
%
%
\begin{figure}
\begin{center}
\includegraphics[width=0.49\textwidth]{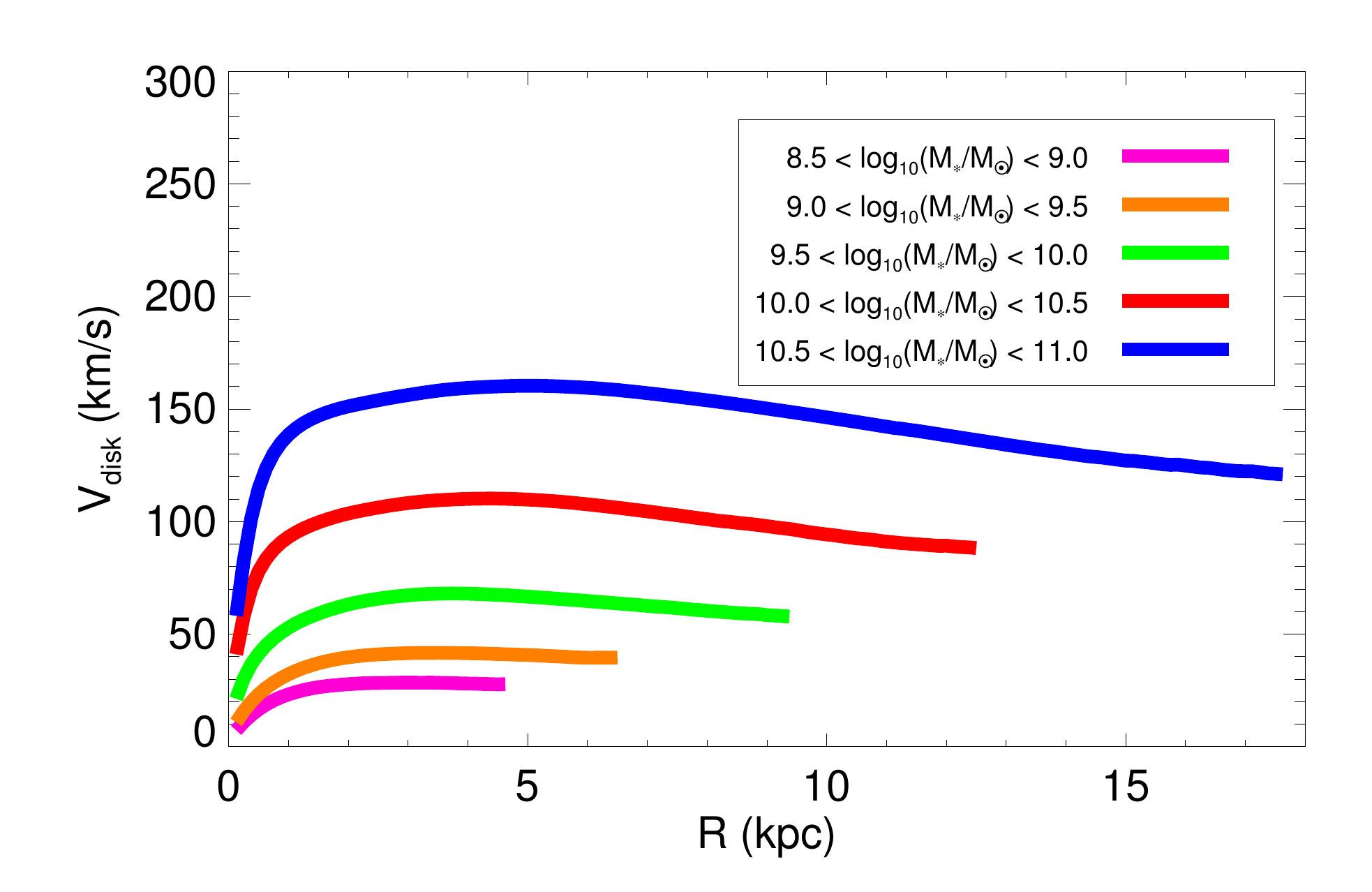}
\includegraphics[width=0.49\textwidth]{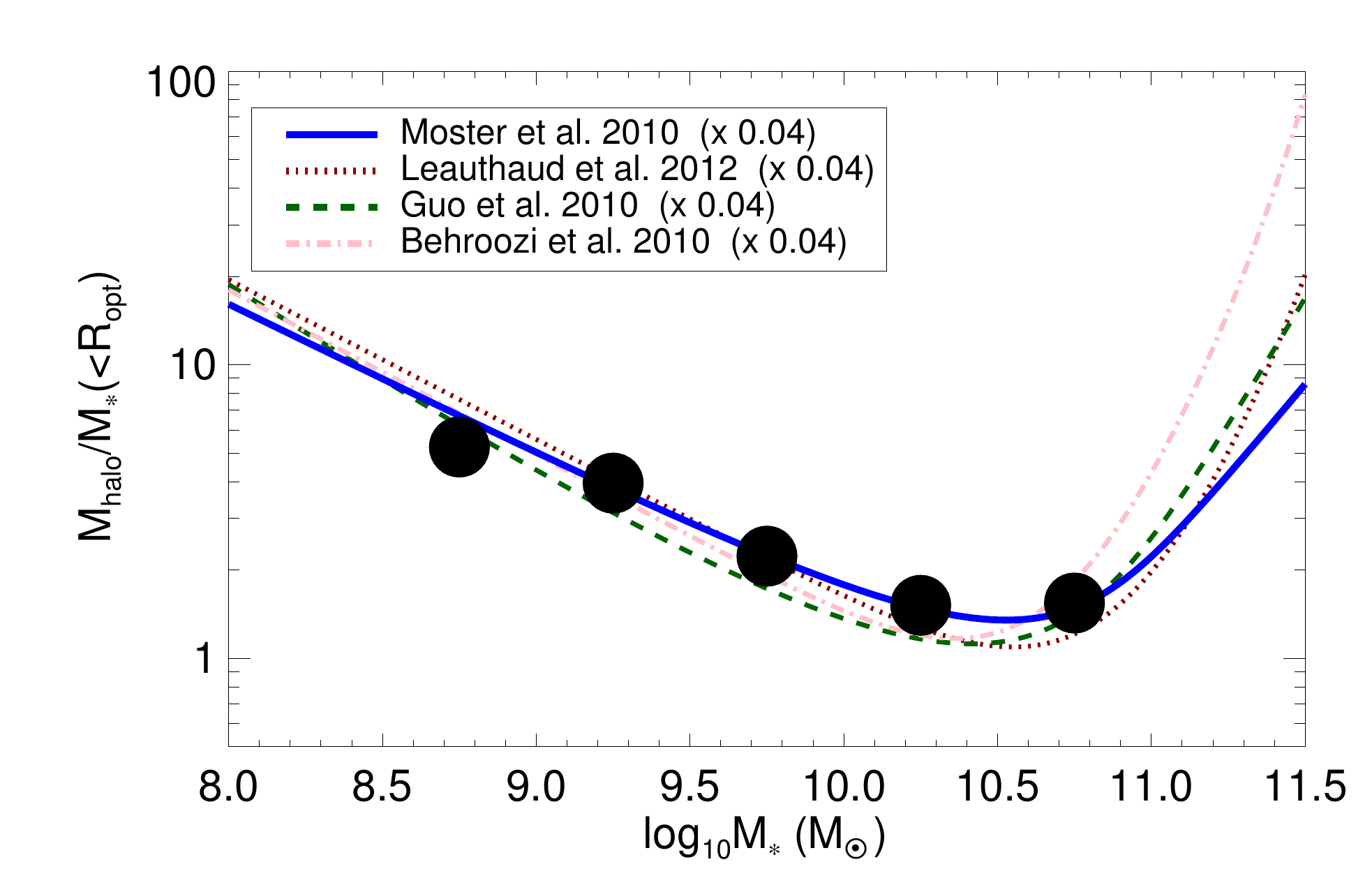}
\vspace{-0.5cm}
\caption{
\emph{Left panel:} 
Mean stellar contribution to the circular velocity for different $M_{\ast}$-bins in the S$^4$G (taken from \cite{DG2016a}). 
\emph{Right panel:} 
The central value of the $M_{\ast}$-bins versus the mean $M_{\rm h}/M_{\ast}(<R_{\rm opt})$ (filled circles). 
As in Fig.~\ref{figuredm}, 
the different lines correspond to estimates in the literature for the \textup{\textup{{\em \textup{total}}}} halo-to-stellar mass ratio vs. the total stellar mass, 
scaled down by a factor 0.04, showing good agreement with our estimate.
}
\label{rotcurve}
\end{center}
\end{figure}
%
%

In the left panel of Fig.~\ref{rotcurve} we show the mean disk(+bulge) component of the rotation curve in bins of $M_{\ast}$, 
obtained by averaging the individual 3.6~$\mu$m rotation curves, rescaled to a common frame in physical units (taken from \cite{DG2016a}). 
Together with the mean stellar density profiles, these curves (and their dispersion) are an observational constraint for galaxy formation models to be checked against. 
From these mean rotation curves, using Eq.~\ref{halo-to-stellar-eq} and the mean H{\sc\,i} velocity amplitudes, 
we inferred the average $M_{\rm h}/M_{\ast}(<R_{\rm opt})$. In the right panel of Fig.~\ref{rotcurve} we reassess 
the relation between the halo-to-stellar mass ratio and $M_{\ast}$, successfully reproducing the slope expected from the $\Lambda$CDM cosmological simulations. 
%
%
\section{Coupling of stellar and dark matter within the disk}
%
%
The reported relation between $M_{\ast}$ and $M_{\rm halo}/M_{\ast}$ means that the present-day galaxies tend to keep in memory the mass of their host (sub)halos, 
where the gas cooled down to form them, even inside the optical radius ($\sim 10\%$ of the halo virial radius). 
We reinforce the idea that dark matter and baryonic matter are intimately coupled in disk galaxies (e.g. \cite{Tully}). 

However, the $M_{\rm h}/M_{\ast}-M_{\ast}$ relation shows a large scatter. Part of this scatter might be due to uncertainties of the H{\sc\,i} line widths, 
which are not the optimal tracer of the halo potential. Partly, it can also arise from the assumption that the gas component is negligible at the optical radius, 
which might not be true for the faintest systems. In individual galaxies, the scatter can be a consequence of a non accurately estimated M/L at 3.6~$\mu$m or 
due to non-stellar contaminants (\cite{Querejeta}). Besides, our definition of $M_{\rm h}/M_{\ast}(<R_{\rm opt})$ can be oversimplified. 
Nonetheless, most probably part of the scatter in the $M_{\rm h}/M_{\ast}-M_{\ast}$ relation also has a physical origin.
Since their birth, galaxy undergo several changes that have an external origin (\cite{kormendy}) 
(e.g. galaxy mergers, gas inflow, or galaxy harassment) that are expected to make the halo-to-stellar mass ratio varying to a certain degree in a cosmic time. 

Using $H_{\alpha}$ kinematics (from \cite{Erroz1}),  Erroz et al. (\cite{Erroz2}) confirmed the scaling relation between the inner slope of the rotation curve and 
the central surface brightness of the galaxy (\cite{Lelli1}). This links the inner density of the potential well to the central stellar density. 
We found that this scaling relation could be recovered by only using the stellar component of the circular velocity, 
confirming that baryonic mass seems to dominate the dynamics in the inner regions (\cite{Lelli2}, \cite{Lelli3}). 
This also manifested in the dependence of the inner velocity gradient (baryons+dark matter) on the bulge-to-total mass ratio and the bulge mass.
%
%
\section{Conclusion}
%
%
We find that the amount of dark matter within the optical disk of S$^4$G galaxies, estimated from the H{\sc\,i} line widths available in the literature, 
scales with the total stellar mass estimated from mid-infrared images, in a manner expected from the $\Lambda$CDM models (e.g. \cite{Moster2010}). 
On average, the amount of dark matter within optical disk is $\sim4\%$ that of the total host halo. 
%
%
\section*{Acknowledgments}   
%
%
We acknowledge financial support to the DAGAL network from the People Programme (Marie Curie Actions) 
of the European Union's Seventh Framework Programme FP7/2007- 2013/ under REA grant agreement number PITN-GA-2011-289313. 
We thank Oulun Yliopiston Akateemiset for a travel grant. 
We thank the organizers of the XII Scientific Meeting of the Spanish Astronomical Society. 
SD thanks SEA for an accommodation grant.
%
%

%
%
\end{document}